\documentstyle[11pt,paspconf,epsf]{article}

\begin{document}

\title{The Andromeda Dwarf Spheroidal Galaxies}
\author{Taft E.\ Armandroff}
\affil{Kitt Peak National Observatory, National Optical Astronomy
Observatories,\altaffilmark{1} P.O.\ Box 26732, Tucson, AZ 85726, USA}
\author{Gary S.\ Da Costa}
\affil{Research School of Astronomy \& Astrophysics, Australian
National University, Private Bag, Weston Creek P.O., ACT 2611,
Australia}
\altaffiltext{1}{The National Optical Astronomy Observatories are
operated by the Association of Universities for Research in
Astronomy under cooperative agreement with the National
Science Foundation.}

\begin{abstract}
Our current knowledge of M31's dwarf spheroidal companions is
reviewed.  Two topics of recent interest constitute the bulk of this
review.  First, color--magnitude diagrams reaching below the horizontal
branch have been constructed for two M31 dwarf spheroidals based on
images from HST/WFPC2.  The horizontal branches are predominantly red
in both galaxies, redder than expected for their metallicity based on
Galactic globular clusters.  Thus, the second parameter effect is seen
in the M31 halo.  Second, recent surveys have revealed three new dwarf
spheroidal companions to M31.  Thus, dwarf spheroidal galaxies are not
as rare around M31 as previously thought and as a result, some
properties of the M31 companion system have changed.
\end{abstract}
\section{Introduction}
Dwarf spheroidal galaxies are probably the most common type of galaxy
in the Universe.  Because of their ubiquity, and because the accretion
of dwarf galaxies may have played an important role in the
formation/evolution of large galaxies, an understanding of dwarf
spheroidal (dSph) galaxies is key to many problems in astronomy.  The
great majority of our knowledge of dwarf spheroidal galaxies results
from the Galaxy's nine dSph companions.  In order to broaden this
knowledge and to investigate the effects of environment on the
formation and evolution of dwarf galaxies, it is important to study
dSph galaxies beyond the Milky Way's companions.  Some mechanisms by
which environment can influence the evolution of dwarf galaxies
include: tidal interactions, removal of a dwarf's interstellar medium
via ram pressure sweeping, confinement of a dwarf's interstellar medium
via ambient pressure, and photoionization of the interstellar medium by
a nearby source of UV radiation (such as a quasar or AGN).

Detailed study of dwarf spheroidals beyond the Milky Way will help
reveal how much we can generalize from the nine Galactic companions.
For example, the Galactic dwarf spheroidals follow reasonably tight
relations between central surface brightness and absolute magnitude and
between absolute magnitude and mean metal abundance (e.g., Caldwell et
al.\ 1992)\@.  Are these relations universal or do they change with
environment?  The present, limited data have not revealed changes with
environment (Caldwell et al.\ 1998)\@.  As another example, the Galactic
dwarf spheroidals show an amazing diversity in their star formation
histories, ranging from a single star-formation episode of globular
cluster age to ongoing star formation that peaked $\sim$3 Gyr ago
(e.g., Da Costa 1998)\@.  Is this diversity universal or did special
circumstances contribute to the varied star formation histories of the
Milky Way's dwarf spheroidals?  Another puzzle is the (imperfect)
correlation among the Galactic dwarfs between the duration of star
formation and distance from the Galactic center, with the outer dwarfs
continuing to form stars over longer periods than the inner dwarfs (van
den Bergh 1994)\@.  Again, it would be valuable to understand whether
this trend is seen in other systems.  Finally,  what is the behavior of
the galaxy luminosity function at the faint end and is this dependent
on environment?  There is some indication that rich clusters of
galaxies have a steeper slope for the faint end of the luminosity
function than the Local Group, though this could be caused by
incompleteness in the Local Group (Trentham 1998).

The nearest collection of dwarf spheroidals beyond the Milky Way's halo
is M31's companions.  Van den Bergh's (1972, 1974) survey of 700 square
degrees around M31 with the Palomar 48-inch Schmidt and IIIaJ plates
revealed three dwarf spheroidal companions (And I, II \& III)\@.
Surface brightness profiles and structural parameters have been
measured for these dSphs by Caldwell et al.\ (1992)\@.
Color--magnitude diagrams from 4m-class ground-based telescopes are
also available for all three galaxies (Mould \& Kristian 1990; Konig et
al.\ 1993; Armandroff et al.\ 1993)\@.  These show the brightest
$\sim$2 magnitudes of the red giant branch and yield line-of-sight
distances similar to that of M31\@.  All these data indicate a
considerable degree of similarity between these three M31 dwarf
spheroidals and those of the Galaxy.  The most significant developments
that have occurred since the last review article devoted exclusively to
the M31 dwarf spheroidals (Armandroff 1994) are:  1) color--magnitude
diagrams from HST/WFPC2 that reach below the horizontal branch for And
I \& II; 2) searches for additional dwarf spheroidals that have
revealed three new M31 companions.  Consequently, this review will
focus on these two areas of recent progress.

\section{WFPC2 Studies}

Although ground-based images of the M31 dSph companions can provide a
large amount of useful information, such data are limited by image
crowding to the brightest 2 or 3 magnitudes of the red giant branch.
On the other hand, because of the Hubble Space Telescope's superior
resolution, images obtained with the WFPC2 camera aboard HST can yield
accurate photometry for stars at the level of the horizontal branch, or
fainter, in these galaxies.  In this section we outline some of the
results obtained in our team's (other Co-I's are N.\ Caldwell and
P.\ Seitzer) program to image the M31 dSph companions with WFPC2\@.  The
results for And~I have been published in Da~Costa et al.\ (1996) while
those for And~II are presented here for the first time.  Our
observations of And~III are scheduled to be executed in early 1999.

\subsection{Observations and Photometry}

Our program uses the $F555W$ (wide--$V$) and $F450W$ (wide--$B$)
filters; the $F450W$ filter being preferred to the more common $F814W$
($\sim$$I$) because it allows greater sensitivity to faint {\it blue}
horizontal branch stars.  For each dSph our procedure is to take a
first series of deep exposures and then, a few ($\leq$5) days later,
take a second series of exposures at the same orientation but with the
field center displaced by $\sim$20 PC pixels.  This displacement of the
field center allows us to readily separate faint stars from the
instrumental effects present in WFPC2 images, while the lag
between the observation sets allows us to detect variable stars,
particularly RR~Lyrae variables.  The total exposure times for each
dSph are of order 2.3 hrs with the $F555W$ filter and 5.4 hrs with
$F450W$.

Even when centered on the centers of these dSph galaxies, the WFPC2
frames are sufficiently uncrowded (cf.\ Fig.\ 1 of Da~Costa et
al.\ 1996) that aperture photometry techniques can be successfully
employed.  We then use the zeropoints and transformation equations
listed in Holtzman et al.\ (1995) to convert the WFPC2 photometry to
the standard $V$, $B-V$ system, though we have a ground-based program
underway to confirm the zeropoints of these data.  The final
color-magnitude diagram (cmd) for And~I derived from our WFPC2 images
is shown in Fig.\ 1, which is taken from Da~Costa et
al.\ (1996).
\begin{figure}[t]
\begin{center}
\begin{minipage}{6cm}
\epsfxsize=8.5cm
\epsfbox{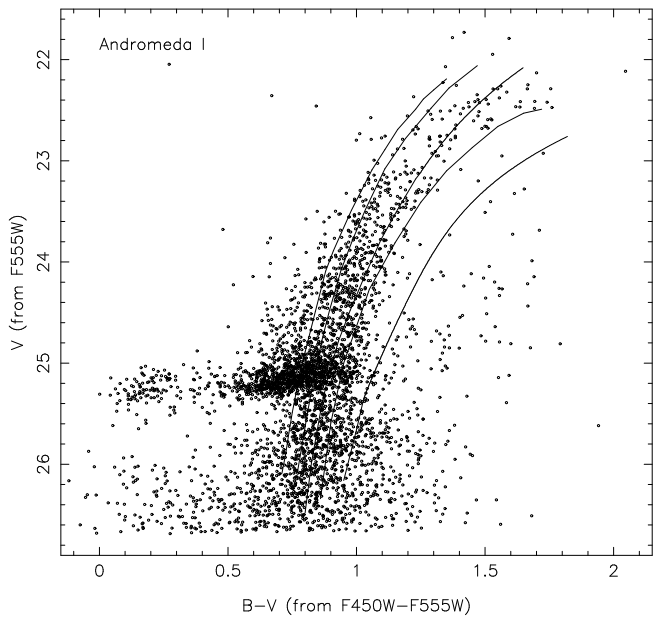}
\end{minipage}
\caption{The $V, (B-V)$ color-magnitude diagram for the M31 dSph
companion And~I from Da~Costa et al.\ (1996)\@.  Note the dominance of
red horizontal branch (HB) stars though blue HB and RR~Lyrae variables
are also present.  The giant branches of the standard Galactic globular
clusters M68 \mbox{([Fe/H]=--2.09)}, M55 (--1.82), NGC~6752 (--1.54),
NGC~362 (--1.28) and 47~Tuc (--0.71) are superposed assuming
$(m-M)_{V}$=24.68 and E($B-V$)=0.04 for And~I, together with an
additional 0.05 mag redward shift to improve the quality of the overall
fit.}
\end{center}
\vspace{-8mm}
\end{figure}

From the data presented in Fig.\ 1, and from the similar
cmd for And~II, we can draw a number of inferences.  We will consider
these in the following order: results that derive primarily from the
morphology of the cmds; results, such as dSph line-of-sight distances,
that depend on the $F555W/V$ zeropoint; and results, such as mean
abundances and abundance dispersions, that depend on the $F555W/V$
zeropoint and the $F450W/B$ transformation and zeropoint.

\subsection{Horizontal Branch Morphologies}

It is immediately apparent from Fig.\ 1 that the morphology of the
horizontal branch (HB) in And~I is dominated by red HB stars.  However,
it is equally obvious that blue HB stars occur and, as Da~Costa et
al.\ (1996) have shown, RR Lyrae variables are also present.  If we
use a HB morphology index $i$ of the form $i = b/(b+r)$, where $b$ and
$r$ are the numbers of blue and red HB stars, respectively (see
Da~Costa et al.\ 1996 for formal definition of these quantities), then
$i$ = 0.13 $\pm$ 0.01 for And~I\@.  As discussed below, the mean
abundance for And~I is $<$[Fe/H]$>$ = --1.45 $\pm$ 0.2 dex.  At this
abundance, the HB morphologies of Galactic globular clusters that
follow the (HB morphology, [Fe/H]) relation defined by the majority of
such clusters, have relatively many more blue HB stars and many fewer
red HB stars than does And I\@.  For example, the standard globular
cluster M5, which has a similar abundance to And~I, has $i$ $\approx$
0.75\@.  For this reason, we can say unequivocally that And~I shows the
second parameter effect in the same way as do many of the Galactic dSph
companions.  Da~Costa et al.\ (1996) go on to argue that in And~I, the
dominance of red HB stars results from a mean age for the bulk of the
population that is younger than that of the Galactic globular clusters,
though the presence of blue HB and RR~Lyrae stars testifies to the
presence of a minority older population.  In other words, the And~I cmd
implies that star formation in this dSph has continued for an
extended period.  This interpretation is supported by the results of
Mighell \& Rich (1996) for the Galactic dSph Leo~II, which has similar
HB morphology and mean abundance to And~I\@.  Mighell \& Rich (1996)
used WFPC2 data that reaches below the main sequence turnoff to
infer that the mean age of the stellar population in Leo~II is $\sim$9
Gyr, with the majority of the star formation having occurred in a
$\sim$4 Gyr interval about that epoch, though some older stars are also
present.

Our WFPC2 cmd for And~II is similar to that shown for And~I in
Fig.\ 1\@.  In particular, once again the HB morphology is dominated by
red stars though, as for And~I, blue HB and candidate RR~Lyrae variable
stars are also present.  Indeed, there are relatively more blue HB
stars in the And~II cmd than is the case for And~I: the HB morphology
index for And~II is $i$ = 0.18 $\pm$ 0.02 (cf.\ $i$ = 0.13 $\pm$ 0.01
for And I)\@.  Given the somewhat higher mean abundance of And~II
($<$[Fe/H]$>$ $\approx$ --1.3, see below) and its large abundance
dispersion, this dSph is less obviously a strong second parameter
candidate in the way And~I is.  Nevertheless, it is likely that the
second parameter effect is at work in And~II\@.  For example, the
standard Galactic globular cluster M4, which has an abundance similar
to the And~II mean, has a relatively evenly populated HB with $i$ =
0.45 $\pm$ 0.05, while NGC~362, which also has an abundance similar to
the And~II mean, has a dominant red HB morphology with $i$ = 0.04 $\pm$
0.02\@.  This latter cluster is one of the `classic' Galactic halo second
parameter clusters and has an age $\sim$2 Gyr younger than the majority
of Galactic halo globular clusters.  We conclude therefore, pending
further analysis, that the And~II cmd does show the second parameter
effect and that the most likely cause is that the bulk of the stellar
population in And~II is somewhat younger than the age of most Galactic
halo globular clusters.

Perhaps the most interesting result concerning HB morphology, however,
is that in And~I it varies with distance from the galaxy's center.
Inside the core radius we find no radial variation in the HB morphology
index, but outside the core radius the relative number of blue HB stars
increases by approximately a factor of two.  In other words, the blue
HB stars in And~I are more widely distributed than the red HB stars
that make up the bulk of the population.  The origin of this radial
gradient is not easily identified but the most likely explanation is
that at the epoch of its initial episode of star formation, the
proto-And~I was more spatially extended than it was when the bulk of
the stars formed.  Apparently similar HB morphology gradients are also
present in two (Sculptor and Leo~II) of the three Galactic dSphs where
there is sufficient data to consider the question.  We have not yet
performed the equivalent analysis on our And~II data.

\subsection{Distances}

For And~I, the mean magnitude of the horizontal branch is $V$ = 25.25
$\pm$ 0.04 mag.  Adopting a reddening E($B-V$) = 0.04 and the distance
scale based on the horizontal branch models of Lee et al.\ (1990), in
which $M_{V}$(HB) = 0.17[Fe/H] + 0.82 mag,\footnote[2]{This relation also
underlies the $I$ magnitude of the red giant branch tip distance scale
(cf.\ Da~Costa \& Armandroff 1990, Lee et al.\ 1993).} the line-of-sight
distance to And~I is 810 $\pm$ 30 kpc for a mean abundance of
$<$[Fe/H]$>$ = --1.45 $\pm$ 0.2 dex.  This distance agrees well with
that, 790 $\pm$ 60 kpc, derived by Mould \& Kristian (1990) from the
$I$ magnitude of the And~I red giant branch tip.  The most directly
comparable distance determinations for M31, based on either red giant
branch tip stars or RR Lyraes in the M31 halo or horizontal branch
stars in M31 globular clusters, are 760 $\pm$ 45 kpc or 850 $\pm$ 20
kpc, both on the same distance scale (see the discussion in Da~Costa et
al.\ 1996)\@.  We can then conclude only that the relative M31/And~I
line-of-sight distance is 0 $\pm$ 70 kpc.  Given that And~I lies
$\sim$45 kpc in projection from the center of M31, the true distance of
And~I from the center of M31 then lies between $\sim$45 and $\sim$85
kpc.  The lower limit is smaller than the galactocentric distances of
any of the Galaxy's dSph companions with the exception of Sagittarius,
while the upper limit is comparable to the galactocentric distances of
the nearer Galactic dSphs such as Ursa Minor, Sculptor and Draco.

For And~II, the mean magnitude of the horizontal branch is $V$ = 24.90
$\pm$ 0.04 mag from which we can immediately infer that And~II lies
closer to us than And~I\@.  Indeed, adopting the same distance scale
and reddening as above, and assuming a mean metal abundance of
$<$[Fe/H]$>$ $\approx$ --1.3 (see next section), we find that And~II
lies at a distance of 690 $\pm$ 25 kpc, or 120 $\pm$ 70 kpc in front of
M31 along the line-of-sight.  This is consistent with the ground-based
results of Konig et al.\ (1993) who found And~II to be 0.4 $\pm$ 0.2
mag closer than M31\@.  On the sky And~II lies closer to M33 than M31.
However, our results now unambiguously associate And~II with M31 since
M33 lies $\sim$150 kpc beyond M31 along the line of sight.  When
combined with And~II's projected distance of $\sim$130 kpc from the
center of M31, these estimates indicate that the true distance of this
dSph from the center of M31 lies between $\sim$150 and $\sim$240 kpc.
The lower limit corresponds approximately to the galactocentric
distance of Fornax while the upper limit is similar to the
galactocentric distances of the Leo systems, the most distant of the
Galactic dSph companions.  And~II, however, may not be the most distant
of the M31 dSph companions, since And VI (see below) lies at least
$\sim$270 kpc (the projected distance) from M31.

\subsection{Mean Metal Abundances and Abundance Dispersions}

With the distances established, the mean metal abundances of these
dSphs can be determined via the comparison of the dSph cmds with the
giant branches of Galactic globular clusters whose [Fe/H] values are
well known.  An example of this is shown in Fig.\ 1 where the giant
branches of five standard globular clusters, whose abundances range
from [Fe/H] = --2.1 to [Fe/H] = --0.7, are overlaid on the And~I WFPC2
observations.  Da~Costa et al.\ (1996) discuss the mean abundance
determination process in some detail and their conclusion is that the
mean abundance of And~I is $<$[Fe/H]$>$ = --1.45 $\pm$ 0.2 dex.  The
uncertainty given includes uncertainty in the calibration relation, the
statistical error in the observed giant branch mean color, and the
effect of an adopted $\pm$0.03 mag systematic uncertainty in the $B-V$
color zeropoint.  This value, however, is in good accord with that,
$<$[Fe/H]$>$ = --1.4 $\pm$ 0.2, determined by Mould \& Kristian (1990)
from the mean $(V-I)_{0}$ color of And~I's upper red giant branch.
Further, there does not appear to be any change in this mean abundance
with distance from the center of And~I.

Application of the same technique to the And~II WFPC2 cmd yields a mean
abundance of $<$[Fe/H]$>$ $\approx$ --1.3, with an uncertainty
comparable to that derived for And~I\@.  This mean abundance is
somewhat more metal-rich than previous ground-based determinations.
For example, Konig et al.\ (1993) give $<$[Fe/H]$>$ = --1.59
(+0.44,--0.12) from their Gunn system $g$,$r$ cmd study while
Armandroff (1994) lists $<$[Fe/H]$>$ = --1.63 $\pm$ 0.25 from the mean
giant branch color in a $(I, V-I)$ cmd.  This relatively high mean
abundance for And~II derived from the WFPC2 data has an interesting
consequence: at $<$[Fe/H]$>$ $\approx$ --1.3 and $M_{V}$ $\approx$
--11.6 (which is based on the new distance derived above), And~II now
deviates significantly from the relation between mean abundance and
absolute magnitude apparently followed by most dSph and dE galaxies
(cf.\ Fig.\ 18 of Caldwell et al.\ 1998)\@.  For example, this mean
abundance for And~II exceeds that of the Fornax dSph, a system which is
more than two magnitudes more luminous.

The lack of image crowding on the WFPC2 images permits the photometric
errors, unlike the case for ground-based images, to approach the photon
statistics limit for all but the very brightest stars, where systematic
effects seem to limit the precision to $\sim$0.015 mag (see Da~Costa et
al.\ 1996)\@.  Consequently, investigating the {\it intrinsic} color
width of the giant branch in these WFPC2 cmds is relatively
straightforward.  The occurrence of internal ranges in
abundance is one of the characteristics of the Galactic dSph galaxies
and the M31 dSph companions prove to be no exceptions.  For And~I, the
results of Da~Costa et al.\ (1996) can be summarized as follows: on the
upper part of the red giant branch, there is an intrinsic color spread
that exceeds by a substantial factor the color spread expected on the
basis of the photometric errors alone.  Applying the same calibration
as for the mean abundance, this color spread can be characterized in a
number of ways.  For example, $\sigma$([Fe/H]) = 0.20 dex, the
inter-quartile range is 0.30 dex, the central two-thirds of the sample
abundance range is 0.45 dex, and the full range of the sample is
$\sim$0.6 dex.  These values are comparable to those seen in Galactic
dSphs.  For example, Suntzeff (1993) lists values of $\sigma$([Fe/H])
for five Galactic dSphs that range from $\sim$0.2 to $\sim$0.3 dex.

On the other hand, Konig et al.\ (1993) have suggested that the
internal abundance range in And~II is unusually large, with
$\sigma$([Fe/H]) $\approx$ 0.43 dex,  though they admit that their
photometric errors, which have to be subtracted in quadrature from their
observed color dispersion, are large and not well constrained.  Thus
the validity of their result is questionable.  However, our WFPC2 data
appear also to indicate that the intrinsic abundance distribution in
this dSph is quite broad, much broader than could be explained by the
photometric errors alone.  Further, because we have essentially two
sets of data resulting from the two different pointings, our
photometric errors are well determined (and small)\@.  Applying a similar
analysis technique to that used by Da~Costa et al.\ (1996) for And~I,
we find that $\sigma$([Fe/H]) $\approx$ 0.38 dex for And~II, while the
inter-quartile range is 0.65 dex, the central two-thirds of the sample
abundance range is 0.85 dex and the full range of the sample is
$\sim$1.4 dex.  These numbers are significantly larger than those for
And~I\@.  We thus have the interesting situation of two M31 companion
dSphs, of similar luminosity and mean abundance, yet whose internal
abundance ranges appear to be substantially different.  This result
reinforces the point that the internal abundance distributions of the
Local Group dSphs apparently do not seem to readily correlate with
luminosity in the way that the mean abundances do.  Indeed the internal
abundance distribution is a characteristic that has the potential to
provide additional significant constraints on the star formation and
enrichment histories in these galaxies.  Further discussion of this
point, however, is beyond the scope of this contribution.

\section{New Dwarf Companions to M31}

\subsection{Searches for Additional M31 Dwarf Spheroidals}

And I, II \& III were found by van den Bergh (1972, 1974) using the
Palomar Schmidt and IIIaJ plates to survey 700 square degrees around
M31\@.  A number of surveys for nearby dwarf galaxies have been
undertaken in recent years.  Irwin (1994) reported a survey for nearby
dwarfs using automated star counts on sky survey plates covering over
20,000 square degrees of high-latitude sky.  This survey revealed the
Sextans dwarf spheroidal (Irwin et al.\ 1990)\@.  Whiting et al.\ (1997)
visually examined survey plates of the entire southern sky, finding the
Antlia dwarf.  Neither of these surveys would be expected to reveal new
M31 companions because, in the first instance, M31 is too distant for
its companions to resolve into stars on sky survey plates and because,
in the second instance, M31 is a northern object.

Because van den Bergh's (1972, 1974) survey yielded a total of three
M31 dwarf spheroidals, when the Galaxy has nine, it has been recognized
that additional M31 companions may be awaiting new surveys. Two other
factors suggest that a new survey may be profitable: 1) the absolute
magnitudes of And I, II \& III are significantly brighter than the
faintest Galactic dwarf spheroidals (see Fig.\ 2 of Armandroff 1994);
2) the radial extent of van den Bergh's (1972, 1974) survey does not
reach the galactocentric distances of the most distant Galactic dwarfs
(see Fig.\ 6 of Armandroff 1994)\@.  Two surveys have recently been
undertaken to find new M31 dwarfs.

The first survey, by Armandroff, Davies \& Jacoby (1998a; see also
Armandroff et al.\ 1999a,b), uses digitized images from the Second
Palomar Sky Survey (POSS-II; Reid et al.\ 1991; Lasker \& Postman
1993)\@.  The POSS-II has better resolution and depth than the POSS-I.
Armandroff et al.\ (1998a) found that cleaning the POSS-II images of
stars and bright galaxies, then applying a matched filter, easily
reveals nearby low surface brightness dwarf galaxies.  They optimized
their filter and detection procedure for the known M31 dwarf
spheroidals. To date, their detection procedure has been applied to
1550 square degrees around M31 (see Fig.\ 2).  Once an object that
resembles the known M31 dSphs on the processed and raw POSS-II images
is found, small telescope CCD imaging is undertaken.  This imaging
eliminates most ``contaminants'' (e.g., distant galaxy clusters,
distant low surface brightness spirals, Galactic cirrus clouds) and
highlights true nearby dwarf galaxies by their incipient resolution
into stars.
\begin{figure}[t]
\caption[]{Map of the region of sky around M31 surveyed for low surface
brightness dwarf galaxies by Armandroff et al.\ (1998a, 1999b), showing
the POSS-II plates that have been searched.  The area surveyed by van
den Bergh (1972, 1974) is outlined in bold.  M31, its known neighbors,
and the new M31 dSphs And V, And VI and Cas are labeled.}
\end{figure}

The second survey, by Karachentsev \& Karachentseva (1999), also uses
the POSS-II and is part of their larger program to search the POSS-II
films for nearby dwarf galaxies (see Karachentseva \& Karachentsev
1998)\@.  They visually searched a circular area of 22$^\circ$ radius
around M31, using morphological criteria to identify nearby low surface
brightness dwarf candidates.

Armandroff et al.\ (1998a,b, 1999a,b) discovered two M31 dwarf
spheroidal candidates: And V \& VI\@.  Karachentsev \& Karachentseva
(1998, 1999) found two strong candidate M31 dwarf galaxies: Pegasus
Dwarf and Cas Dwarf. One candidate was found independently by both
groups: And VI = Pegasus Dwarf.  The Cas Dwarf is located in a region
that lacked POSS-II digital data; therefore it was not found by
Armandroff et al.\ (see Fig.\ 2).

\subsection{Properties of the Newly Discovered Dwarf Galaxies}

Three of the candidate M31 dwarf spheroidals found by Armandroff et
al.\ and Karachentsev \& Karachentseva have been resolved into stars,
indicating that they are indeed nearby dwarf galaxies.  In this
section, we summarize what is known about these three galaxies as of
November 1998.

Armandroff et al.\ (1998a) presented follow-up observations of And V
from the KPNO 4-m telescope and prime-focus CCD imager in the $V$, $R$,
$I$ and H$\alpha$ narrow-band filters.  In the broad-band filters, And
V resolves nicely into stars and exhibits a smooth stellar
distribution, resembling the other M31 dwarf spheroidals.  And V does
not exhibit the features of classical dwarf irregulars, such as obvious
regions of star formation or substantial asymmetries in its stellar
distribution.  In the And V continuum-subtracted H$\alpha$ image, no
diffuse H$\alpha$ emission or H {\sc ii} regions are detected.  The
lack of H$\alpha$ emission in And V reinforces the conclusion, based on
And V's appearance on the broad-band images, that it is a dwarf
spheroidal galaxy rather than a dwarf irregular.  And V is not detected
in any of the IRAS far-infrared bands either.  Because far-infrared
emission traces warm dust, and because some Local Group dwarf irregular
galaxies are detected by IRAS, And V's lack of far-infrared emission
serves as additional, weaker evidence that it is a dSph.

And V's apparent central surface brightness was measured by Armandroff
et al.\ (1998a): 25.7 mag/arcsec$^2$ in $V$\@.  And V has a fainter
apparent central surface brightness than And I, II \& III (24.9, 24.8,
and 25.3 mag/arcsec$^2$ in $V$, respectively; Caldwell et
al.\ 1992)\@.  And V probably eluded detection until now due to its
very dim apparent surface brightness.  Armandroff et al.\ (1998a) also
constructed a color--magnitude diagram for And V stars, in order to
determine its distance and stellar populations characteristics.
Color--magnitude diagrams for the parts of the images dominated by And
V stars reveal a red giant branch, which is absent in the outer regions
of the images.  The tip of the red giant branch is well defined in the
cmd and in the luminosity function.  A distance was derived for And V
based on the $I$ magnitude of the tip of the red giant branch (Da Costa
\& Armandroff 1990, Lee et al.\ 1993)\@. On the distance scale of Lee
et al.\ (1990), the resulting And V distance is 810 $\pm$ 45 kpc. As
discussed in Sec.\ 2.3, the best and most comparable estimates of the
M31 distance are 760 $\pm$ 45 kpc or 850 $\pm$ 20 kpc. This implies
that And V is located at the same distance along the line of sight as
M31 to within the uncertainties.  And V's projected distance from the
center of M31 is 112 kpc; And I, II \& III have projected M31-centric
distances of 46, 144 and 69 kpc, respectively.  The above line-of-sight
and projected distances strongly suggest that And V is indeed
associated with M31.

Armandroff et al.\ (1998a) compared And V's color--magnitude diagram
with fiducials representing the red giant branches of Galactic globular
clusters that span a range of metal abundance (Da Costa \& Armandroff
1990)\@. Based on the position of the And V giant branch relative to
the fiducials, the mean metal abundance of And V is
$\approx$ --1.5, which is normal for a dSph (e.g., Fig.\ 9 of
Armandroff et al.\ 1993)\@.  No bright blue stars are present in the
And V cmd.  Interpreting via isochrones, this lack of blue stars rules
out any stars younger than 200 Myr in And V and is further evidence
that And V is a dwarf spheroidal and not a dwarf irregular.  From the
luminosities and numbers of upper asymptotic giant branch stars in a
metal-poor stellar system, one can infer the age and strength of its
intermediate age component (Renzini \& Buzzoni 1986)\@.  Using the And
V field-subtracted luminosity function, there is no evidence for upper
asymptotic giant branch stars that are more luminous than and redward
of the red giant branch tip.  Therefore, And V does not have a
prominent intermediate age population; in this sense, it is similar to
And I \& III.

And VI = Pegasus Dwarf has been resolved into stars and studied by two
groups.  Armandroff et al.\ (1998b) imaged And VI with the KPNO 4-m
telescope prime-focus CCD on 1998 January 23 in $V$ for 300 seconds.
And VI resolved nicely into stars in this short $V$ image, suggesting
that it is indeed a nearby dwarf galaxy.  On 1998 July 15, Armandroff
et al.\ (1999b) obtained deeper And VI images with the KPNO 4-m
telescope through the $R$ and H$\alpha$ filters.  The $R$ image of And
VI exhibits a smooth stellar distribution and shows a resemblance with
the other M31 dwarf spheroidals.  And VI does not look lumpy or show
obvious regions of star formation, suggesting that it is a dwarf
spheroidal as opposed to a dwarf irregular.  In the And VI
continuum-subtracted H$\alpha$ image, no diffuse H$\alpha$ emission or
H {\sc ii} regions are detected.  The lack of H$\alpha$ emission rules
out current high-mass star formation in And VI and serves as further
evidence that And VI is a dwarf spheroidal.  Like And I, II, III \& V,
And VI is not detected in any of the IRAS far-infrared bands.
Armandroff et al.\ (1999b) are constructing a cmd for And VI based on
images obtained with the WIYN 3.5-m telescope during excellent seeing
conditions ($B$, $V$ \& $I$).

Grebel \& Guhathakurta (1999) have presented a cmd for And VI based on
images from the Keck-II telescope in $V$ and $I$, which clearly shows
the red giant branch.  Based on the $I$ magnitude of the red giant
branch tip, they determined a line-of-sight distance of 780 $\pm$ 40
kpc, associating And VI with M31.  From the position of the red giant
branch relative to standard globular cluster fiducials, Grebel \&
Guhathakurta found $<$[Fe/H]$>$ = --1.2 $\pm$ 0.3.  They also measured
an intrinsic central surface brightness of 24.5 mag/arcsec$^2$ in $V$.

Grebel \& Guhathakurta (1999) also obtained similar data for the Cas
Dwarf from Keck-II.  As for And VI, their cmd clearly shows the red
giant branch, and from it they derived a distance and mean
metallicity.  Their line-of-sight distance of 710 $\pm$ 35 kpc places
the Cas Dwarf in the extended M31 satellite system.  The resulting
metallicity is $<$[Fe/H]$>$ = --1.3 $\pm$ 0.3.  Grebel \& Guhathakurta
also measured an extinction-corrected central surface brightness of
23.6 mag/arcsec$^2$ in $V$\@.  Both And VI and the Cas Dwarf have
apparent central surface brightnesses that are much brighter than And
V, and somewhat brighter than And I, II \& III.  Both And VI and the
Cas Dwarf are outside of van den Bergh's (1972, 1974) survey area,
which likely accounts for their anonymity until now.

Once accurate $M_V$ values are measured for And V, And VI and Cas, and
once their $<$[Fe/H]$>$ values are established with definity, it will be
important to re-evaluate whether the M31 and Galactic dSphs follow the
same relations between absolute magnitude, central surface brightness,
and mean metal abundance.  The sample will be twice as large and the
data of higher quality than the previous determinations (e.g., Fig.\ 7
of Armandroff 1994).

\subsection{Impact}

The discovery of the M31 dwarf spheroidals And V, And VI = Pegasus
Dwarf, and Cas Dwarf doubles the number of known M31 dSphs.  This
changes the properties of M31's satellite system, as discussed below.
The most obvious change is that M31 is not as poor in dwarf spheroidals
as previously thought.

Karachentsev (1996) discussed the spatial distribution of the
companions to M31.  The discovery of And V, And VI and Cas changes the
spatial distribution of the M31 satellites.  Curiously, And I, II \&
III are all located south of M31, while the three more luminous dwarf
elliptical companions NGC 147, 185 \& 205 are all positioned north of
M31.  Also, Karachentsev (1996) noted that there are more M31
companions overall south of M31 than north of M31.  The locations of
And V and Cas north of M31 lessen both of these asymmetries (though And
VI is south of M31; see Fig.\ 2).  With projected radii from the center
of M31 of 112, 271 and 224 kpc, respectively, And V, And VI and Cas
increase the mean projected radius of the M31 dwarf spheroidals from 86
kpc to 144 kpc.

The discovery of nearby dwarf galaxies like And V, And VI and Cas
augments the faint end of the luminosity function of the Local Group.
We do not yet have reliable $M_V$ values for And V, And VI or Cas, but
they appear to be in the range $-12 < M_V \leq -10$.  This increases the
number of galaxies in the Local Group in this $M_V$ range by 27\%.
From a survey of nine clusters of galaxies, Trentham (1998) derived a
composite luminosity function that is steeper at the faint end than
that of the Local Group (see his Fig.\ 2)\@.  He attributed the
difference to poor counting statistics and/or incompleteness among the
Local Group sample.  The discovery of And V, And VI and Cas reduces
somewhat the discrepancy between the Local Group luminosity function
and the extrapolation of Trentham's (1998) function.

\section{Future Directions}
Several opportunities for advancing our understanding of M31's dwarf
spheroidal companions are apparent:
\begin {itemize}
\item The search for dSph companions to M31 is not yet complete.  The
ongoing searches based on the POSS-II may reveal additional
companions.  The Sloan Digital Sky Survey should allow a significantly
deeper search for faint, low surface brightness M31 companions.
\item It will be valuable to obtain HST/WFPC2 cmds for all six M31
dwarf spheroidals.  These cmds will yield precise line-of-sight distances,
hopefully facilitating a 3-dimensional map of the distribution of M31
satellites.  Whether this distribution is random or exhibits
``streams'' will provide an interesting comparison with the Galactic
dSph distribution.  Also, having HB morphologies for all six known M31
dSphs will allow a more comprehensive evaluation of the second
parameter effect in the halo of M31.
\item No information is currently available about the H {\sc i} content
of And II, And V, And VI or Cas Dwarf via 21 cm observations.  Either H
{\sc i} detections or strict upper limits would be valuable.  The upper
limits for And I \& III should be refined (see Thuan \& Martin 1979).
\item Measurements of radial velocity are possible for red giants at
the M31 distance with 8--10 meter telescopes.  One can constrain the
total mass of M31 via the velocity dispersion of its dwarf satellite
system (as, for example, Zaritsky et al.\ 1989 did for the Galaxy).
The impressive projected radii of And VI and Cas will result in mass
limits to very large radii.
\item Radial velocities will also allow mass-to-light ratios (M/L) to
be derived for the M31 dSphs.  Measurement of a high M/L for an M31
dSph would be a step toward universalizing the large velocity
dispersions and consequent large M/L values observed for Galactic
dSphs.  Also, the large galactocentric radii of And II, And VI and Cas
should render tidal effects negligible.
\end{itemize}

\acknowledgments
We are grateful to our collaborators, Nelson Caldwell and Pat Seitzer
on the And II WFPC2 cmd, and James Davies and George Jacoby on the
M31 dSph survey, for allowing us to present results in advance of
publication and for helpful discussions.

\end{document}